%% file: grmhd_dynamo.tex
\documentclass[trackchanges,twocolumn,resetfootnote]{aastex701}
\usepackage[T1]{fontenc}
\usepackage{graphicx}
\usepackage{amsmath}
\usepackage{amssymb}
\usepackage{bm}
\usepackage{soul}
\usepackage{hyperref}
\hypersetup{
  colorlinks   = true,
  citecolor   = blue
}
\usepackage{natbib}
\usepackage{color}

\input{macros}

\begin{document}

\title{Subgrid Mean-field Dynamo Model with Dynamical Quenching in General Relativistic Magnetohydrodynamic Simulations}

\author[orcid=0000-0002-2991-5306]{Hongzhe Zhou}
\email[show]{hzzhou@sspu.edu.cn}
\affiliation{School of Mathematics, Physics and Statistics, Shanghai Polytechnic University, 2360 Jinhai Road, Shanghai, 201209, China}
\affiliation{Tsung-Dao Lee Institute, Shanghai Jiao Tong University, Shanghai, 201210, China}

\author[orcid=0000-0002-8131-6730]{Yosuke Mizuno}
\email[show]{mizuno@sjtu.edu.cn}
\affiliation{Tsung-Dao Lee Institute, Shanghai Jiao Tong University, Shanghai, 201210, China}
\affiliation{School of Physics and Astronomy, Shanghai Jiao Tong University, Shanghai, 200240, China}
\affiliation{Key Laboratory for Particle Physics, Astrophysics and Cosmology (MOE), Shanghai Key Laboratory for Particle Physics and Cosmology, Shanghai Jiao-Tong University, Shanghai, 200240, China}

\author[orcid=0000-0001-9189-860X]{Zhenyu Zhu}
\email[show]{zhenyu.zhu@rit.edu}
\affiliation{Center for Computational Relativity and Gravitation, Rochester Institute of Technology, Rochester, NY 14623, USA}
\affiliation{Tsung-Dao Lee Institute, Shanghai Jiao Tong University, Shanghai, 201210, China}

\begin{abstract}

Large-scale magnetic fields are relevant for a number of dynamical processes in accretion disks, including driving turbulence, reconnection events, and launching outflows.
Numerical simulations have indicated that the initial strengths and configurations of the large-scale magnetic fields have a direct imprint on the outcome of an accretion disk evolution.
To facilitate future self-consistent simulations that include intrinsic dynamo processes, we derive and implement a subgrid model of a helical large-scale dynamo with dynamical quenching in general-relativistic resistive magnetohydrodynamical simulations of geometrically thin accretion disks.
By incorporating previous numerical and analytical results of helical dynamos,
our model features only one input parameter, the viscosity parameter $\alphaSS$.
We demonstrate that our model can reproduce butterfly diagrams seen in previous local and global simulations.
With rather aggressive parameter choice of $\alphaSS=0.02$ and black hole spin $\aBH=0.9375$, our thin-disk model launches weak collimated polar outflows with Lorentz factor $\simeq 1.2$, but no polar outflow is present with less vigorous turbulence or less positive $\aBH$.
With negative $\aBH$, we find the field configurations to appear more similar to Newtonian cases,
whereas for positive $\aBH$, the poloidal field loops become distorted and the cycle period becomes sporadic or even disappears.
Moreover, we demonstrate how $\alphaSS$ can avoid to be prescribed and instead be determined by the local plasma beta.
Such a fully dynamical subgrid dynamo allows for self-consistent amplification of the large-scale magnetic fields.

\end{abstract}


\section{Introduction}

Magnetic fields are ubiquitous in accretion systems, including protoplanetary disks, X-ray binaries, tidal disruption events (TDEs), quasi-periodic eruptions, as well as active galactic nuclei \citep[e.g.,][]{EHT2021VII,Liodakis+2023,You+2023,EHT2024VII,Nicholl+2024,Pasham+2024,Ohashi+2025}.
Prominent roles played by magnetic fields in accretion disks include
driving the magneto-rotational instability (MRI),
transporting angular momentum directly or indirectly through MRI,
driving disk wind,
as well as launching jets \citep[e.g.,][]{BlandfordZnajek1977,BalbusHawley1991,BaiStone2013}.
All these processes require magnetic fields to be coherent at scales greater than the turbulence scale.

The origin of such large-scale magnetic fields remains elusive.
Advecting magnetic field lines from outer disk regions is efficient only for geometrically thick disks \citep{Cao2011,Ressler+2020,Dhang+2023},
whereas for geometrically thin disks, \textit{in situ} amplification of magnetic fields, i.e., dynamos, is likely the dominant mechanism.
Magnetization of thin disks is crucial for understanding the high-soft state of X-ray binaries, protoplanetary disks, and radio-quiet AGNs.
It may also be relevant for understanding radio-loud AGNs,
which may have puffy centers but cool and thin outer disks.

Magnetic dynamos in accretion flows have been typically studied in local shearing-box simulations,
in addition to a number of more costly Newtonian global disk simulations \citep[e.g.,][]{Gressel2010,GresselPessah2015,HoggReynolds2018,Dhang+2020,GresselPessah2022,Dhang+2024MRI,Reboul-Salze+2025}.
Based on the typical scales of the amplified magnetic fields, dynamos can be characterized into the small-scale and the large-scale types.
Both types require a developed turbulent flow, which poses a difficulty in realizing dynamos in general-relativistic magnetohydrodynamics (GRMHD) simulations.
In order to have an \textit{ab initio} dynamo process, simulations have to start with a weak seed field, which in turn requires a high resolution to resolve the MRI and a long time evolution to reach a saturated state.
Such constraints lead to only one existing GRMHD simulation that has a large-scale dynamo,
\cite{Liska+2020} \citep[and for its analysis,][]{Jacquimin-Ide+2024}, although its initial magnetic field is still mildly strong.

A subgrid dynamo model provides an alternative approach to realizing dynamical large-scale magnetic fields in GRMHD simulations.
In such models, subgrid transport coefficients are constructed to represent the effects from unresolved turbulence, including dynamo-driving terms and turbulent diffusion,
and have been implemented in a number of GRMHD simulations of accretion disks \citep{vonRekowski+2003,BucciantiniDelZanna2013,Bugli+2014,StepanovsFendt2014,Skadowski+2015,FendtGassmann2018,Dyda+2018,Tomei+2020,VourellisFendt2021,Shibata+2021},
and simulations of binary neutron star mergers
\citep[e.g.,][]{Giacomazzo+2015,Most2023}
However, a shortcoming of the existing implementation is that the spatial profile of the transport coefficients has remained prescribed instead of depending on other physical quantities such as density, sound speed, or the amplified magnetic field itself.
In particular, as the amplified field becomes dynamically important, its feedback onto the dynamo (known as dynamo quenching) has not been properly modeled.
As a result, the strength and the morphology of the saturated large-scale magnetic fields remain dependent on either
(i) a phenomenological quenching formula with a prescribed saturation level, or
(ii) the numerical diffusivity to balance the dynamo term, which itself depends on the numerical scheme and resolution.

In fact, the profile and the evolution of the transport coefficients have been studied with a number of methods in various types of simulations.
Such attempts have been done in Newtonian shearing-box simulations that apply to geometrically thin disks \citep[e.g.,][]{Brandenburg+2008,Gressel2010,GresselPessah2015,Dhang+2024MRI},
and is recently extended to global disk simulations or GRMHD simulations of geometrically thick disks \citep{Dhang+2020,Jacquimin-Ide+2024,Duez+2025}.
The discovered transport coefficients have been used to build a nonlinear helical dynamo model with dynamical quenching in a Newtonian disk by \cite{Zhou2024}.

In this work, we extend the model of \cite{Zhou2024} to resistive GRMHD simulations.
The formulation of the mean-field dynamo is based on the proposed closure of \cite{BucciantiniDelZanna2013}, which is a natural extension of the Newtonian case. Furthermore, in our model, the dynamo coefficients are no longer free parameters \citep[as in][]{BucciantiniDelZanna2013} but expressed in terms of other macroscopic quantities, as done in \cite{Zhou2024}.
We also modify the Newtonian dynamical quenching formalism to be compatible with resistive GRMHD.

The rest of this work is organized as follows.
In Section~\ref{sec:dynamo}, we generalize the subgrid dynamo formulation with dynamical quenching to GRMHD cases, and introduce our numerical setup.
In Section~\ref{sec:fiducial}, we present the results from a fiducial run and discuss its accretion dynamics and dynamo properties.
In Section~\ref{sec:param}, we explore the parameter survey and investigate the consequences of varying the viscous parameter and black hole spin.
We also propose a fully dynamical model in which the only dynamo model parameter, namely the viscous parameter, is allowed to vary with the magnetization.
We discuss possible applications and relevant implications of our model,
and conclude in Section~\ref{sec:conclusion}.

\section{GRMHD mean-field dynamo with dynamical quenching}
\label{sec:dynamo}

\subsection{A brief review of dynamical quenching in Newtonian MHD}

The framework of large-scale (or mean-field) dynamos \citep{Parker1955,Steenbeck+1966} has been used to understand the amplification and saturation of magnetic fields over temporal-spatial scales greater than those of the turbulence.
As the large-scale field becomes dynamically strong, its back-reaction onto the turbulent flow gradually drives the system to a saturated state, a process is known as dynamo quenching.

Dynamos possessing kinetic and magnetic helicity are expected to operate in stratified and rotating flows, such as geometrically thin accretion disks.
For such helical dynamos, the quenching process is well understood and is referred to as the dynamical quenching \citep{Pouquet+1976,BlackmanField2002,BrandenburgSubramanian2005}.
In the rest of this subsection, we briefly outline the idea in the Newtonian case.

We start from Ohm's law,
\beq
\bm E+\bm V\times\bm B=\eta_0\bm J,
\eeq
where $\bm E$, $\bm V$, $\bm B$, and $\bm J$ are the electric field, the velocity, the magnetic field, and the current density, respectively, and $\eta_0$ is the microscopic resistivity.
By taking an ensemble average on both sides and moving to a co-moving frame where $\bar{\bm V}=\bm 0$, we obtain the mean-field Ohm's law,
\beq
\bar{\bm E}+\bar{\bm v'\times\bm b'}=\eta_0\bar{\bm J},
\eeq
where the mean fields are denoted with overlines,
and the primed lower cases are the residual turbulent fields,
e.g., $\bm v'=\bm V-\bar{\bm V}$ etc.
The cross terms like $\bar{\bar{\bm V}\times\bm b'}$ and $\bar{\bm v'\times\bar{\bm B}}$ vanish due to Reynolds' rules of averaging.
Combining with mean-field Faraday's law $\partial_t\bar{\bm B}=-\del\times\bar{\bm E}$, we obtain the mean-field induction equation,
\beq
\partial_t \bar{\bm B}=\del\times\left(\bar{\bm v'\times\bm b'}-\eta_0\bar{\bm J}\right),
\eeq
and $\bm\emf=\bar{\bm v'\times\bm b'}$ is the mean-field electromotive force (EMF).
To find a closure, the EMF is expanded in terms of the spatial gradients of $\bar{\bm B}$.
For isotropic turbulence, the ansatz is
\beq
\bar{\bm v'\times\bm b'}=-\xi\bar{\bm B}-\eta\bar{\bm J},
\eeq
where $\xi$ is the dynamo driver\footnote{%
$\xi$ is typically denoted as $-\alpha$ in dynamo literatures.},
and $\eta=l^2/3\tau$ is the turbulent resistivity with $l$ and $\tau$ being the turbulence length and time scales, respectively.
Our goal of building a subgrid dynamo model is to express $\xi$ and $\eta$ using mean-field quantities.

We separate the dynamo-driving coefficient $\xi$ into a kinematic and a dynamical component, $\xi=\xik+\xid$,
where $\xik$ comes from the underlying MHD turbulence when the large-scale magnetic field is dynamically weak, 
and $\xid$ includes the back-reaction of the large-scale field when it is dynamically strong.
Such a formalism is slightly different from the conventional ``kinetic+magnetic'' description \citep[e.g.,][]{BlackmanField2002},
where the kinetic and the magnetic components are from the turbulent velocity field and the turbulent magnetic field, respectively.
We prefer the ``kinematic+dynamical'' convention because for MRI, the background turbulence is necessarily nonlinear and $\xi$ already includes (or even is dominated by) the turbulent magnetic field when the large-scale field is weak.
The $\xid$ term then includes those contributions caused by the dynamically strong large-scale field.

Convection-driven turbulence in stratified rotators possesses a net kinetic helicity $\propto\bm\Omega\cdot\del\ln\bar{\rho v'^2}$ \citep{Steenbeck+1966,ZhouBlackman2019},
where $\bm\Omega$ is the rotation vector and $\bar\rho$ is the mean density.
Helical dynamos in such a turbulent flow simply have $\xik\propto\bm\Omega\cdot\del\ln\bar{\rho v'^2}$, being negative in the northern hemisphere and positive in the southern hemisphere.
For MRI turbulence, numerically inferred profiles of $\xik$ typically have an additional sign reversal near the equator caused by buoyant small-scale magnetic flux ropes \citep{BrandenburgDonner1997,Brandenburg1998,Gressel2010,HoggReynolds2018,GresselPessah2015,Dhang+2020,Dhang+2024MRI}.

The kinematic term $\xik$ drives helical large-scale field,
and the conservation of the total magnetic helicity requires a build-up of the small-scale helical fields.
This results in a dynamical contribution $\xid$ that is proportional to the small-scale current helicity density \citep{Pouquet+1976,BlackmanField2002},
\beq
\xid=-\frac{\tau}{3\bar\rho}\bar{\bm j'\cdot\bm b'}.
\eeq
In each hemisphere, $\xid$ always has an opposite sign to the local $\xik$.
For example, a positive $\xik$ drives negatively helical large-scale magnetic fields, which are necessarily accompanied by positively helical small-scale fields, and hence $\xid$ is negative.

\subsection{Implementation in a Newtonian thin disk}

To implement $\xik$ and $\xid$ as subgrid terms in numerical simulations,
it is necessary to express the unresolved turbulence scales $l$ and $\tau$, and the current helicity density $\bar{\bm j'\cdot\bm b'}$ in terms of large-scale quantities \citep{Zhou2024,Duez+2025}.
To estimate $l$ and $\tau$, we use the classical relation of the turbulent viscosity $\nu=l^2/3\tau=\alphaSS\cs \hh$ \citep{ShakuraSunyaev1973} (where $\cs$ is the sound speed,
$\hh$ is the density scale height,
and $\alphaSS$ is the viscosity parameter that quantifies turbulence vigorousness),
along with $\tau=\Omega^{-1}$ and the vertical hydrostatic balance $\cs=\Omega \hh$, to get
\beq
l\simeq\sqrt{3\alphaSS}\hh,\ \tau\simeq\Omega^{-1}.
\eeq
This also results in a turbulent velocity scale $v'=l/\tau\simeq \sqrt{3\alphaSS}\cs$.

For rotating convection-driven turbulence,
the kinematic term $\xik'$ has the form \citep{Steenbeck+1966,ZhouBlackman2019}
\beq
\xik'\simeq l^2\bm\Omega\cdot\del\ln\bar{\rho v'^2}.
\eeq
For a Gaussian vertical profile $\bar\rho\propto e^{-z^2/2\hh^2}$ and an ideal-gas equation of state $p\propto\rho^{\hat\gamma}$ (where $p$ is the gas pressure and $\hat\gamma$ is the adiabatic index),
we find $\xik'=-\hat\gamma\alphaSS\Omega z$ \citep{Zhou2024}.
The specific sign reversal for MRI turbulence near the equator is incorporated by introducing an empirical factor, and the general expression for the kinematic term $\xik$ is written as
\begin{align}
\xik
=&\tanh\left(\left|\frac{z}{\hh}\right|-2\right)\xik'\notag\\
=&-\hat\gamma\tanh\left(\left|\frac{z}{\hh}\right|-2\right)
\alphaSS\Omega z.
\end{align}

To find an expression for the dynamical dynamo coefficient $\xid$, we assume a local conservation of magnetic helicity\footnote{%
For a covariant formulation of helicity transport, see \cite{WuMost2024}.},
\beq
\bar{\bm A}\cdot\bar{\bm B}+\bar{\bm a'\cdot\bm b'}=0,
\eeq
where $\bar{\bm A}$ and $\bm a'$ are the mean and the turbulent magnetic vector potential, respectively.
Furthermore, using $l$ and $\hh$ as the scales of the small- and the large-scale fields, we have
\beq
\bar{\bm j'\cdot\bm b'}\simeq l^{-2}\bar{\bm a'\cdot\bm b'}
\simeq -l^{-2}\bar{\bm A}\cdot\bar{\bm B}
\simeq -\frac{1}{3\alphaSS}\bar{\bm J}\cdot\bar{\bm B}.
\label{eqn:helicity_balance}
\eeq
Thus the form of $\xi$ has been closed.

Shearing-box simulations have revealed that MRI turbulence has comparable turbulent viscosity and turbulent resistivity \citep{FromangStone2009,GuanGammie2009,LesurLongaretti2009},
so that $\eta\simeq\nu=\alphaSS\Omega\hh^2$.
For MRI turbulence, we supplement $\eta$ with a vertical profile extracted from shearing boxes \citep{Gressel2010,GresselPessah2015,Zhou2024}
\beq
\eta=\frac{3}{8}\left(1+\frac{z^2}{\hh^2}\right)\alphaSS\Omega\hh^2.
\eeq

Finally, we take into account the decrease in the sound speed (and hence the turbulence length scale $l$) when moving away from the disk mid-plane.
Using the ideal-gas equation of state,
$\cs^2\propto \rho^{(\hat\gamma-1)/2}$ with $\hat\gamma=5/3$,
one finds $l,\cs\propto e^{-z^2/6\hh^2}$.
The full dynamo prescription is then
\begin{align}
&\xi=\xik
+\frac{1}{9\alphaSS\rho\Omega}\bar{\bm J}\cdot\bar{\bm B},
\label{eqn:xitot_newtonian}\\
&\xik=-\hat\gamma
\tanh\left(\left|\frac{z}{\hh}\right|-2\right)
e^{-z^2/6\hh^2}\alphaSS \Omega z,
\label{eqn:xik}\\
&\eta=\frac{3}{8}\left(1+\frac{z^2}{h^2}\right)e^{-z^2/3\hh^2}\alphaSS\Omega\hh^2.
\label{eqn:eta}
\end{align}

Since Newtonian MHD neglects the displacement current and $\bar{\bm E}$ is not dynamical,
one has $\bar{\bm J}=\del\times\bar{\bm B}/\mu_0$ with $\mu_0$ being the vacuum magnetic permeability.
The relation closes the equations, and Equation~(\ref{eqn:xitot_newtonian}) is sufficient to be implemented in simulations, as is done in \cite{Zhou2024}.
However, in relativistic MHD, both the electric and the magnetic fields are dynamical, and $\xid$ should depend on both.
To facilitate a direct comparison between the Newtonian and the relativistic cases, we rewrite Equation~(\ref{eqn:xitot_newtonian}) in an equivalent form using the mean-field Ohm's law.
Substituting $\bar{\bm J}=\left(\bar{\bm E}-\xi\bar{\bm B}\right)/\left(\eta+\eta_0\right)$, we find
\beq
\xi=\frac{\xik+\varpi\bar{\bm E}\cdot\bar{\bm B}}
{1+\varpi B^2},
\label{eqn:quenching_newtonian}
\eeq
where $\varpi^{-1}=9\alphaSS(\eta+\eta_0)\rho\Omega=9\alphaSS p(\eta+\eta_0)/\Omega\hh^2\sim\mathcal{O}(9\alphaSS^2p)$.

We note that, coincidentally, the denominator of Equation~(\ref{eqn:quenching_newtonian}) is similar to a commonly used quenching formula, $\xi=\xik/\left(1+B^2/B_\text{sat}^2\right)$,
where $B_\text{sat}$ is some prescribed saturation value.
However, Equation~(\ref{eqn:quenching_newtonian}) cannot be approximated by $\xi=\xik/\left(1+\varpi B^2\right)$ and the value of $\varpi^{-1}$ does not play a role in determining the saturation strength.
The correct quenching process comes from the second term in the numerator, which offsets $\xik$, i.e., physically, the quenching happens due to the balance between the small-scale kinetic and current helicity.

In what follows, we generalize the dynamical quenching formalism to the general relativity case.
From now on, we drop the overlines, and all the quantities shall be regarded as mean fields, unless otherwise specified.

\subsection{Dynamo prescription in GRMHD}

We perform numerical simulations of accretion flows around Kerr black holes using the publicly available Black Hole Accretion Code \citep[BHAC;][]{Porth+2017,Olivares+2019} with its resistive GRMHD revision \citep{Ripperda+2019BHAC}.
In the 3+1 formalism, Maxwell's equations are
\begin{align}
&\partial_t\left(\gamma^{1/2}E^j\right)
+\partial_i\left[\gamma^{1/2}\left(
\beta^jE^i-\beta^iE^j
\right.\right.\notag\\
&\left.\left.\qquad\quad
-\gamma^{-1/2}\eta^{jik}\alpha B_k\right)\right]
=-\gamma^{1/2}\left(\alpha J^j-q\beta^j\right),
\label{eqn:E}\\
&\partial_t\left(\gamma^{1/2}B^j\right)
+\partial_i\left[\gamma^{1/2}\left(
\beta^jB^i-\beta^iB^j
\right.\right.\notag\\
&\left.\left.\qquad\quad
+\gamma^{-1/2}\eta^{jik}\alpha E_k\right)\right]
=0,\\
&\gamma^{-1/2}\partial_i\left(\gamma^{1/2}E^i\right)=q,\\
&\gamma^{-1/2}\partial_i\left(\gamma^{1/2}B^i\right)=0,
\end{align}
where $E^i$, $B^i$, $J^i$, and $q$ are the electric field, magnetic field, current density, and charge density measured by an Eulerian observer,
$\gamma=\sqrt{-g}/\alpha$ is the determinant of the spatial $3$-metric $\gamma_{ij}$,
$\alpha$ and $\beta^i$ are the lapse function and the spatial shift vector, respectively,
and $\eta^{ijk}$ is the Levi-Civita symbol.

We consider axisymmetric two-dimensional simulations, for which the solved physical quantities shall be interpreted as representing azimuthally averaged mean fields from a three-dimensional counterpart.
The residual turbulent fields are regarded as unresolved or poorly resolved in our simulations, and their effects on the evolving physical quantities shall be included using subgrid dynamo terms.

Our mean-field dynamo closure is based on the proposal of \cite{BucciantiniDelZanna2013},
in which Ohm's law is modified from $e^\mu=\eta_0 j^\mu$ to
\beq
e^\mu=\xi b^\mu+(\eta+\eta_0) j^\mu,
\label{eqn:Ohm's law}
\eeq
where $e^\mu$, $b^\mu$, and $j^\mu$ are the electric field, magnetic field, and current density, respectively, measured in the comoving frame of the fluid,
$\eta_0$ is the microscopic resistivity,
and $\xi$ and $\eta$ are dynamo coefficients to be specified.
The $\xi$ term in Equation~(\ref{eqn:Ohm's law}) is responsible for generating  poloidal and toroidal magnetic fields from each other,
while differential rotation only generates the toroidal component from the poloidal ones.
Hence, the dynamo model that we construct is of an $\alpha^2\Omega$ type.
However, we have verified that only retaining a finite $\xi$ in the $e^\phi$ equation (i.e., an $\alpha\Omega$ type dynamo) has a minor influence on the result, because the toroidal EMF is dominated by the $\Omega$ effect.

The 3+1 decomposition of Equation~(\ref{eqn:Ohm's law}) gives the current density
\begin{align}
J^i
=&qv^i +\frac{\Gamma}{\eta+\eta_0}\left(
E^i+\Gamma^{-1/2}\eta^{ijk}v_jB_k-v^iv^jE_j\right)\notag\\
&-\xi\frac{\Gamma}{\eta+\eta_0}\left(B^i-\Gamma^{-1/2}\eta^{ijk}v_jE_k-v^iv^jB_j\right),
\label{eqn:J}
\end{align}
where $\Gamma=(1-v^2)^{-1/2}$ is the Lorentz factor.
The last term in Equation~(\ref{eqn:J}) corresponds to the dynamo effect from the unresolved turbulence, and the turbulent resistivity $\eta$ reflects the diffusion of large-scale magnetic fields to unresolved turbulent ones.

The turbulent resistivity is roughly $\Rm$ (the magnetic Reynolds number) times larger than the microscopic resistivity, but still remains numerically small and renders the last two terms in Equation~(\ref{eqn:J}) stiff.
An implicit-explicit (IMEX) scheme is typically used to resolve this issue
\citep{BucciantiniDelZanna2013},
including the resistive GRMHD version of \textsc{BHAC} \citep{Ripperda+2019BHAC}, in which the conservative-to-primitive procedure and the Ohm's law are solved together in the implicit step.
In principle, the newly added dynamo term in Equation~(\ref{eqn:J}) modifies this step as well, making the problem more complicated \citep{Tomei+2020}.
We simplify the problem by noting that the dynamo coefficients $\xi$ and $\eta$ reflect only the average properties of the underlying turbulence, which evolve on time scales longer than the local orbital time scale and therefore much longer than one time step.
Hence, it becomes a good approximation to replace the values of $\eta$ and the entire last term in Equation~(\ref{eqn:J}) by those calculated from the last time step.
In our practice, the last term in Equation~(\ref{eqn:J}) is treated as a source term in Equation~(\ref{eqn:E}), and the rest is solved using the existing IMEX scheme in \textsc{BHAC}.

The Newtonian expressions~(\ref{eqn:xik}) and (\ref{eqn:eta}) naturally carry over to the general relativity case.
To generalize Equation~(\ref{eqn:xitot_newtonian}), we promote the magnetic field and the current density to be the comoving fields $b^\mu$ and $j^\mu$.
Using the mean-field Ohm's law~(\ref{eqn:Ohm's law}) to solve for $j^\mu$, we arrive at
\beq
\xi=
\frac{\xik+\betacrit/\beta_{eb}}
{1+\betacrit/\betamag},
\label{eqn:quenching_gr}
\eeq
where $\beta_{eb}\equiv 2p/e^\mu b_\mu$, $\betamag\equiv 2p/b^\mu b_\mu$ is the plasma beta calculated from the large-scale magnetic field, and $\betacrit\equiv2\Omega\hh^2/9\alphaSS(\eta+\eta_0)$.
Equation~(\ref{eqn:quenching_gr}) is natural generalization of the Newtonian case~(\ref{eqn:quenching_newtonian}).

The dynamo action is restricted within the disk region using a number of criteria:
(i) The density is greater than the floor density profile;
(ii) The plasma beta is finite and greater than unity,
and the magnetization $\sigma=b^\mu b_\mu/\rho$ is between zero and unity;
(iii) $|z|<4\hh$;
(iv) $\cos\theta<0.8$ and $r>r_\text{ISCO}$, where $r_\text{ISCO}$ is the radius of the inner-most stable orbit (ISCO).
Here, conditions (iii) and (iv) are geometrical, and the latter is used to exclude the jet and the plunging regions.

We note that our determination of the active region of the dynamo is not yet fully dynamical, and the prescription of the dynamo coefficients relies on an analytical steady-state solution.
In particular, the disk mid-plane should coincide with $z=0$ in order to validate Equation~(\ref{eqn:xik}).
Hence, our subgrid model does not apply to disks with time-changing structures, e.g., precessing disks or disks in TDEs.

\subsection{Numerical setup}

We set up axisymmetric two-dimensional thin-disk GRMHD simulations following the prescription of \cite{Dihingia+2021}.
The initial condition is prescribed in Boyer-Lindquist (BL) coordinates,
although the evolving equations will be solved in the Modified Kerr-Schild (MKS) geometry \citep{Porth+2017}.
We use a unit system where $G=M=c=1$.
The disk initially has a Gaussian vertical density profile $\propto e^{-z^2/2\hh^2}$ and Keplerian orbital frequency $\Omega=1/\left(\aBH+r_\text{BL}^{3/2}\right)$ (where $\aBH$ is the black hole spin),
and the scale height is parameterized using $\hh=\sqrt{\Theta_0/8}r_\text{BL}$ with $\Theta_0$ being a given constant reference temperature\footnote{%
The form of $\hh$ is chosen to match the definition of $\Theta_0$ in \cite{Dihingia+2021}.}.
We also use an adiabatic index of $\hat\gamma=5/3$.
For a more detailed description of the initial disk profile,
we refer the readers to \cite{NovikovThorne1973}, \cite{PageThorne1974}, and \cite{Dihingia+2021}.

We use $3$ levels of static mesh refinement, reaching an effective resolution of $512\times256$ in the inner equatorial plane.
The initial density maximum occurs at $r_\text{BL}=200$, and the simulation domain covers $r_\text{BL}\leq 1500$ and $\theta_\text{BL}\in[0,\pi]$.
The density floor is set to be $\rho_\text{fl}=10^{-5}r_\text{BL}^{-3/2}$,
and the pressure floor is $p_\text{fl}=10^{-7}r_\text{BL}^{-5/2}$.
We set a ceiling for the Lorentz factor of $\Gamma_\text{max}=20$.
We initialize the magnetic field using the vector potential with $A_{r,\theta}=0$ and
\begin{align}
A_\phi\propto
&\max\left(0,\rho-0.01\rho_\text{max}\right)\notag\\
&\times\cos\theta_\text{BL}
\sin\frac{\pi(r_\text{BL}-r_\text{BL,in})}{3},
\label{eqn:init_A}
\end{align}
and the initial density-averaged $\betamag>10^{6}$ is dynamically weak.
The microscopic resistivity is set to be $\eta_0=10^{-8}$ in code units,
and is always much smaller than the turbulent resistivity.

To ensure that the disk remains thin over long-time evolution, we add an artificial cooling following the prescription of \cite{Noble+2009}.
More specifically, we employ an isotropic cooling function
\beq
\Lambda=\Omega\rho\epsilon\sqrt{2Y-2}
\eeq
wherever $Y=p/\rho\Omega^2\hh^2>1$ and the gas is bounded,
where $\epsilon=p/(\hat\gamma-1)$ is the specific internal energy density.
The cooling source terms are added to the equations of the covariance three-momentum and the rescaled energy density; see Equations~(1) and (2) in \cite{Dihingia+2023}.

Numerical simulations and observations of quiescent-phase X-ray binaries and AGNs typically find $\alphaSS\lesssim0.02$ \citep{Starling+2004,Fromang+2007,Hawley+2011,Flock+2012,HameuryLasota2016,Duez+2025} \citep[protoplanetary disks have smaller values,][]{Rafikov2017}.
We use fiducial values of $\alphaSS=0.02$ and $\Theta_0=0.1$ in \run{A1} (resulting in dimensionless scale height $\sqrt{\Theta_0/8}\simeq 0.11$),
and \run{A0} is its non-dynamo counterpart with $\alphaSS=0$.
In group~\texttt{B}, we vary the black hole spin.
In group~\texttt{C}, we change the value of $\alphaSS$,
including runs~\texttt{C3} and \texttt{C4} that employ a varying $\alphaSS$ which depends on $\betamag$ (see Section~\ref{sec:dynamical}).
A summary of the runs is shown in Table~\ref{tab:runs}.

\begin{table}[htp]
\caption{Parameters of the simulation runs.
The saturated plasma beta $\beta_\text{sat}$ is defined as $1/\left(\abra{\betamag^{-1}}_\rho\right)$ averaged over a $4\times 10^4\tg$ time interval at the end of each run.
The saturated normalized flux $\phi_\text{sat}$ is averaged in the same time interval.}
\begin{center}
\input{runs.tex}
\end{center}
\label{tab:runs}
\end{table}

\subsection{Diagnostics}

We use angle brackets, $\abra{Q}$, to denote the volume average of a quantity $Q$.
A density-weighted average is defined as $\abra{Q}_\rho=\abra{\rho Q}/\abra{\rho}$.
Furthermore, the normalized magnetic flux across the horizon is defined as $\phi=\Phi/{\dot M}^{1/2}$, where $\Phi$ is the magnetic flux across the horizon and $\dot M$ is the accretion rate.

\section{The fiducial run}
\label{sec:fiducial}

\subsection{Plasma dynamics}

\begin{figure*}
\centering
\includegraphics[width=0.9\textwidth]{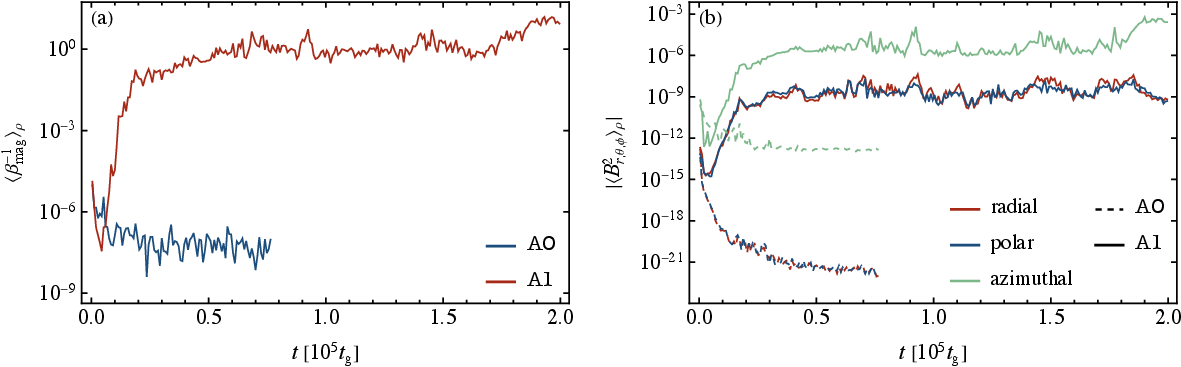}
\caption{The time series of the density-weighted averages of
(a) the inverse plasma beta, and (b) the squared magnetic strengths for the non-dynamo \run{A0} and the fiducial dynamo \run{A1}.
}
\label{fig:A1_ts_A}
\end{figure*}

We first present the results of the fiducial run, \texttt{A1}, and compare with a non-dynamo \run{A0} whose $\alphaSS=0$, i.e., without any subgrid dynamo or turbulent diffusion term.
Figure~\ref{fig:A1_ts_A} shows the time series of the density-weighted average values of the inverse plasma beta and the strengths of the three components of the magnetic field for these two runs.
While \run{A0} only exhibits a marginal decrease in the overall magnetization,
\run{A1} becomes more strongly magnetized thanks to the subgrid dynamo effect.
In the saturated state, the density-averaged value of $\betamag$ for \run{A1} reaches $\mathcal{O}(1)$ values, reflecting the formation of a mildly magnetized disk.
The saturated energy of the toroidal magnetic field is about $10^3$ times that of the poloidal components, which is one to two orders of magnitude larger than that reported by \cite{Jacquimin-Ide+2024}, but notice that the latter is based on a geometrically thick disk:
With sub-Keplerian rotation, the $\Omega$ effect that generates the azimuthal field component becomes weaker.

\begin{figure*}
\centering
\includegraphics[width=0.9\textwidth]{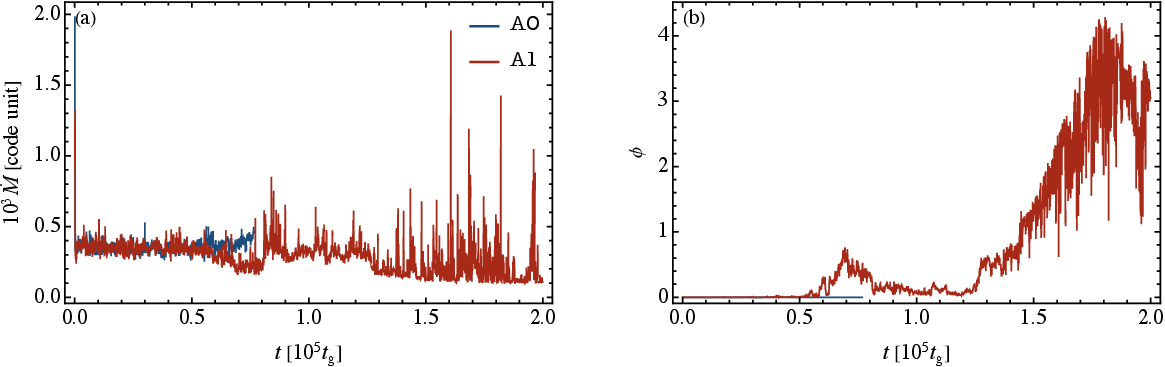}
\caption{Same as Fig.~\ref{fig:A1_ts_A} but shown the the time series of
(a) accretion rate at the horizon, and
(b) normalized magnetic flux at the horizon.
}
\label{fig:A1_tsH_A}
\end{figure*}

\begin{figure*}
\centering
\includegraphics[width=0.8\textwidth]{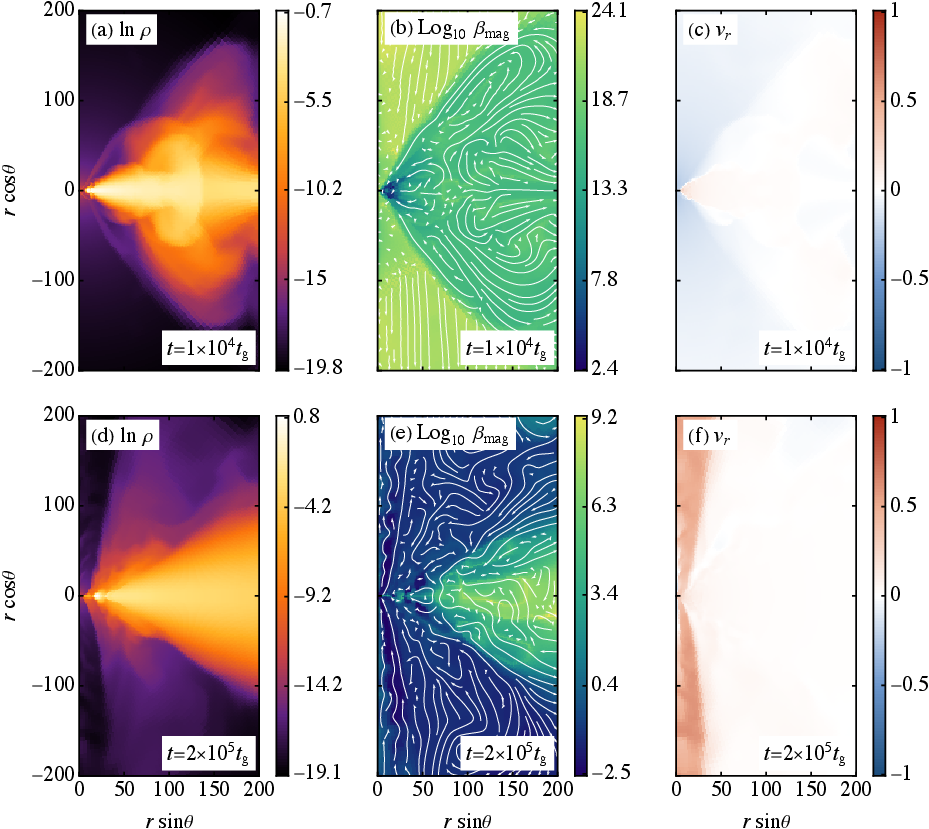}
\caption{Snapshots of \run{A1}
for the distribution of the density (left), 
the plasma beta with magnetic field lines (middle),
and the radial velocity (right),
at $t=10^4\tg$ (top) and $t=2\times 10^5\tg$ (bottom).
}
\label{fig:A1_snaps}
\end{figure*}

Regarding accretion dynamics, Figure~\ref{fig:A1_tsH_A}(a) reveals that there is only a minor increase in the mean accretion rate for \run{A1} due to the dynamo-amplified magnetic field, but significant variability occurs.
At the end of the simulation, the density-weighted values of the MRI quality factors are $Q_{r,\theta}<0.1$.
Here the quality factors are defined as the ratios between the fastest growing MRI mode in each direction and the grid resolution.
For example, $Q_\theta=\lambda_\theta/\Delta x_\theta$ with
\beq
\lambda_\theta=\frac{2\pi}{\sqrt{\left(\rho h+b^2\right)\Omega}}b^\mu e_\mu^{(\theta)}
\eeq
and $\Delta x_\theta=\Delta x^\mu e_\mu^{(\theta)}$ being calculated in the tetrad basis $e_\mu^{(\hat\alpha)}$ of the fluid frame.
Increasing the resolution may help resolve the vertical modes better.
We note that having intrinsic MRI (e.g., by increasing the resolution or having thicker disks) will not invalidate our subgrid model, because the large-scale fields driven by MRI will also contribute to the dynamo quenching and shut the dynamo down at self-consistent strengths.
Despite of the minor change in the accretion rate,
the normalized magnetic flux at the horizon increases significantly as seen in Figure~\ref{fig:A1_tsH_A}(b).
The resulting $\phi\simeq 3$ is below the threshold of magnetically arrested regime,
but strong enough to drive polar outflows [see Figure~\ref{fig:A1_snaps}(f)].

Since our simulation is axisymmetric, no azimuthal mode is present; therefore, the MRI, if any, remains in the standard weak-field regime.
Previous local shearing-box simulations have indicated that the resulting $\alphaSS$ is on the order of $\mathcal{O}(10^{-2})$ \citep[e.g.,][]{Simon+2012}, self-consistent with the value we set.
We discuss a model where the profile of $\alphaSS$ is self-consistently determined by the local value of plasma beta in Section~\ref{sec:dynamical}.

Our formalism can be naturally extended to 3D cases, where MRI with azimuthal modes is present.
In such cases, suprathermal toroidal magnetic fields can be generated starting from $\betamag\lesssim10^2$ at the initial state \citep{Salvesen+2016,Mishra+2020}, the latter possibly as an outcome of an axisymmetric large-scale dynamo.
It is therefore possible that a magnetically elevated disk can be produced from some weak field with $\betamag\gtrsim 10^6$ in a future 3D implementation.

We plot the distribution of density, the plasma beta (along with the magnetic field lines),
and the radial velocity in Figure~\ref{fig:A1_snaps}, for two snapshots in the early kinematic dynamo phase (top row) and the later saturated dynamo phase (bottom row), respectively.
In the latter stage, low-beta coronal regions form below and above the disk,
penetrated by large-scale magnetic fields that rise from the disk.
The accreted magnetic flux at the horizon can drive weak bipolar outflows [panel (f)] with $\sigma=b^2/\rho<1$ and a Lorentz factor of $\simeq1.2$,
similar to the least magnetized case in \cite{Dihingia+2021}.

\subsection{Dynamo properties}

\begin{figure*}
\centering
\includegraphics[width=0.9\textwidth]{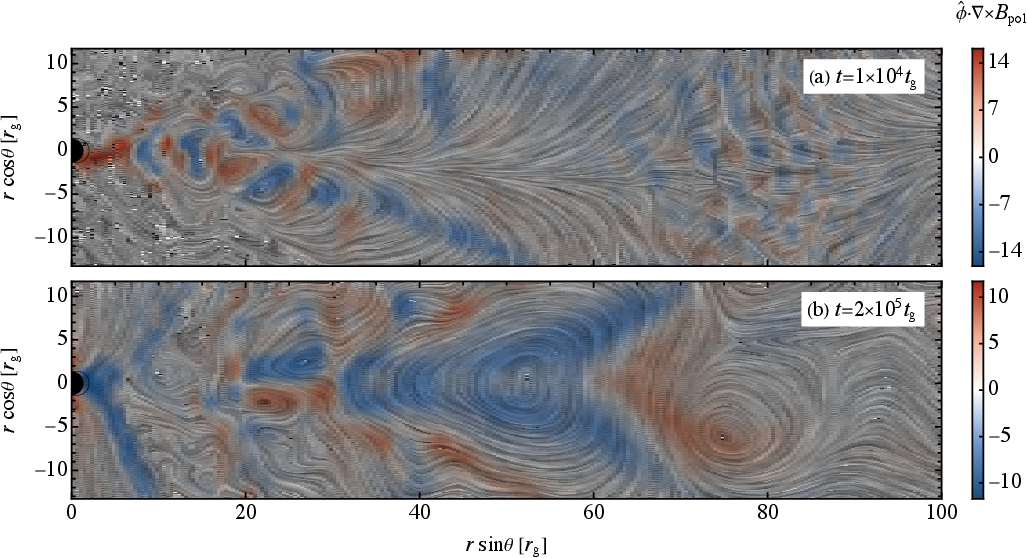}
\caption{For \run{A1}, distribution of line-integral convolution of the poloidal magnetic field $\bm B_\text{pol}$ at two representative snapshots ($t=10^4$ and $2\times10^5t_g$).
Colors indicate the values calculated from the curl of the fields in the code unit to reflect the field orientation.
For each panel, the black dot at $r=0$ indicates the outer horizon, and the thin black curve denotes the ergosphere.
}
\label{fig:A1_Bpol}
\end{figure*}

We next investigate the field polarities due to the dynamo
and examine how the dynamo quenching process works.
The polarities of the dynamo-amplified fields are plotted in Figure~\ref{fig:A1_Bpol}, where we show the line-integral convolution of the poloidal magnetic field in the kinematic (top) and the saturated (bottom) phases, respectively.
The preferred length scale with maximal growth rate is expected to be $\eta/|\xi|\simeq \hh$, and indeed we see that large-scale fields can be maintained against turbulent diffusion.
At saturation, field reversals are frequently seen in the disk, resembling the multiple magnetic loop configuration that some previous works have explored \citep{Nathanail+2020,Jiang+2023,Nathanail+2025}.
Once accreted into the inner disk region, the reconnection of these field lines may lead to variability of emission and flares \citep{Jiang+2024, Jiang+2025}.
We note that the current work has employed a geometrically thin disk whose dynamo mechanisms are better understood, as opposed to the thick disks discussed in these multi-loop simulations.
It remains to be demonstrated that similar field morphologies can be produced through large-scale dynamos in the latter cases.

\begin{figure*}
\centering
\includegraphics[width=0.95\textwidth]{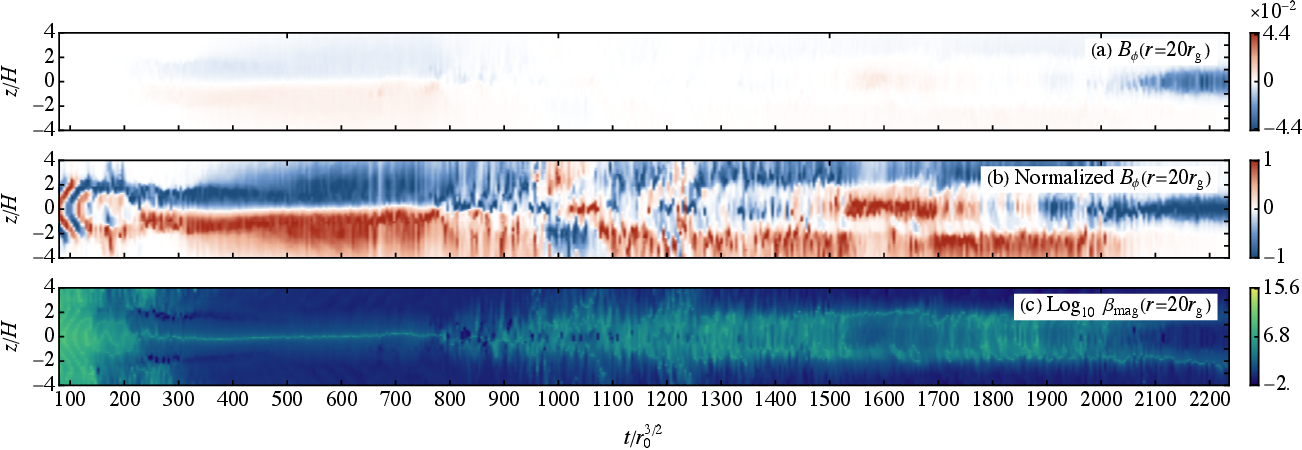}
\caption{For \run{A1}, the space-time diagram at $r=20\rg$ for (a) the azimuthal magnetic field $B_\phi$, (b) $B_\phi$ normalized by its maximal amplitude at each fixed time (to reveal the polarity), and (c) the plasma beta.
}
\label{fig:A1_butterfly}
\end{figure*}

In Figure~\ref{fig:A1_butterfly}, we show the space-time diagrams of $B_\phi$, its polarity, and the plasma beta at $r=r_0=20\rg$.
The saturated dynamo phase approximately starts at $t\simeq 200r_0^{3/2}\simeq 2\times10^4\tg$,
characterized by the onset of mildly strong magnetization.
In panel (b), a butterfly diagram is seen in the kinematic stage at $t/r_0^{3/2}\lesssim 150$, similar to those observed by \cite{Bugli+2014}.
The cyclic pattern soon ends, followed by a relatively long quasi-steady phase in the interval $t/r_0^{3/2}\in[300,900]$,
and then transits to be quasi-periodic again.
Noticeably, in the second butterfly phase, the patterns have extended to higher latitude $|z|\gtrsim2\hh$ and the dynamo period becomes longer.
The field continues to evolve into a quasi-steady phase again when $t/r_0^{3/2}\gtrsim 1500$.
To summarize, the dynamo seems to transit between quasi-periodic and quasi-steady states, which have not been reported in previous local or global simulations.
Such behavior could be an outcome of the coupling between the GR effects and the dynamo in the inner disk region,
possibly combined with a competition between the advected field and the local dynamo-amplified field.
In Section~\ref{sec:aBH}, we show that the persistence of a quasi-periodic dynamo is also associated with the black hole spin, an evidence supporting a GR-intercepted dynamo period.

\begin{figure}
\centering
\includegraphics[width=0.5\textwidth]{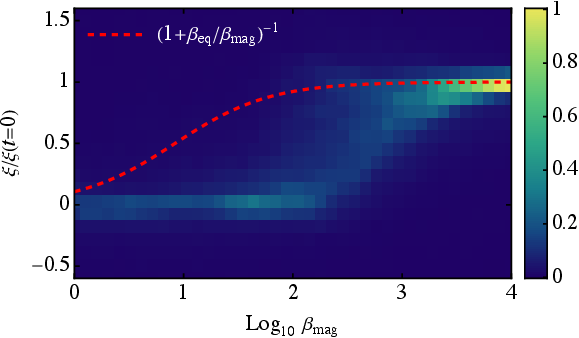}
\caption{For \run{A1}, $\xi$ normalized by its initial value versus $\betamag$ for all the dynamo-active locations and time.
Color indicates the normalized number density of mesh points.
The dashed red curve indicates the algebraic quenching prescription.}
\label{fig:A1_xi_betamag}
\end{figure}

As the magnetization becomes strong,
the dynamo driver $\xi$ is quenched according to Equation~(\ref{eqn:quenching_gr}).
Notice that $\xi$ needs to not vanish at a fully quenched state, but only be suppressed to a value just balancing the turbulent diffusion.
We plot the dynamo coefficient $\xi$ at all dynamo-active mesh points normalized by its initial value against the magnetization $\betamag$ in Figure~\ref{fig:A1_xi_betamag}.
For reference, we also plot the common prescription $\xi=\xi_0(1+\beta_\text{eq}/\betamag)^{-1}$ (usually referred to as the algebraic quenching) as the red dashed curve,
where the energy equipartition value is $\beta_\text{eq}=2p/\rho v^2\simeq 1/6\alphaSS$.
We observe that the algebraic formula leads to a weaker quenching effect and cannot fairly approximate the data points.
With our helicity-constrained quenching, $\xi$ is almost fully quenched at $\betamag\simeq10^2$.

%
%

\section{Comparative study}
\label{sec:param}

In this section, we base on the fiducial \run{A1} and vary the black hole spin $\aBH$ and the viscous parameter $\alphaSS$ to observe the consequences on plasma dynamics and dynamo properties.
Here, $\aBH$ is physical and determined by the initial condition, but $\alphaSS$ is artificial and initially introduced by \cite{ShakuraSunyaev1973} to parameterize turbulent viscosity.
In the last part of this Section, we will relax this last artificial parameter $\alphaSS$ and aim to build a fully dynamical subgrid model.

\subsection{Varying black hole spin}
\label{sec:aBH}

The black hole spin can have several effects on the disk dynamo and the plasma dynamics:
(i) Close to the event horizon, the frame-dragging effect modifies the differential rotation profile and hence the dynamo $\Omega$ effect.
Retrograde flows have a steeper rotation curve compared to prograde ones for the same black hole spin, and a stronger $\Omega$ effect is expected for the former.
(ii) A less rapidly rotating black hole has a slightly weaker gravitomagnetic dynamo effect \citep{KhannaCamenzind1994}, but in our case, this is expected to be subdominant compared to the subgrid $\alpha$ effect.
(iii) For the same amount of horizon-threading magnetic flux, the jet power is lower for lower $\aBH$ \citep{BlandfordZnajek1977,Tchekhovskoy+2010,Narayan+2022}.

\begin{figure*}
\centering
\includegraphics[width=0.9\textwidth]{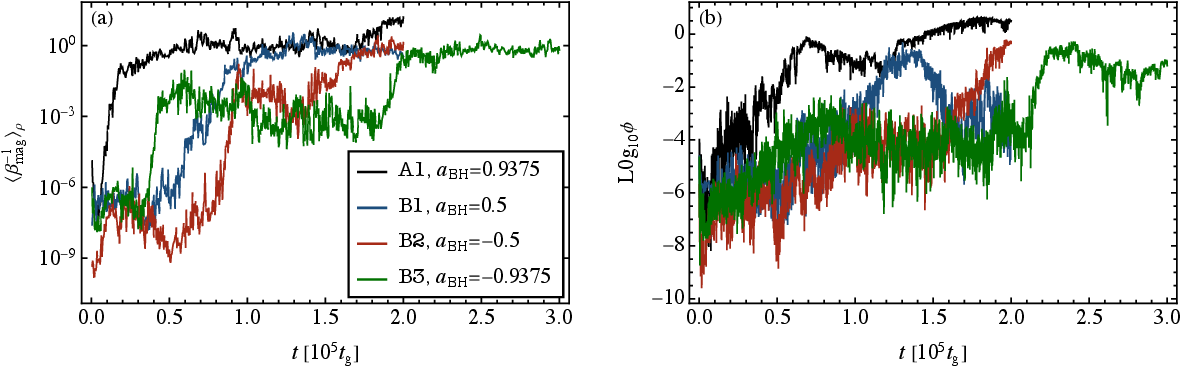}
\caption{With varying $a_\text{BH}$,
the time series of 
(a) the density-weighted averages of $\betamag^{-1}$, and
(b) the normalized magnetic flux at the horizon.}
\label{fig:B12_ts}
\end{figure*}

\begin{figure*}
\centering
\includegraphics[width=1\textwidth]{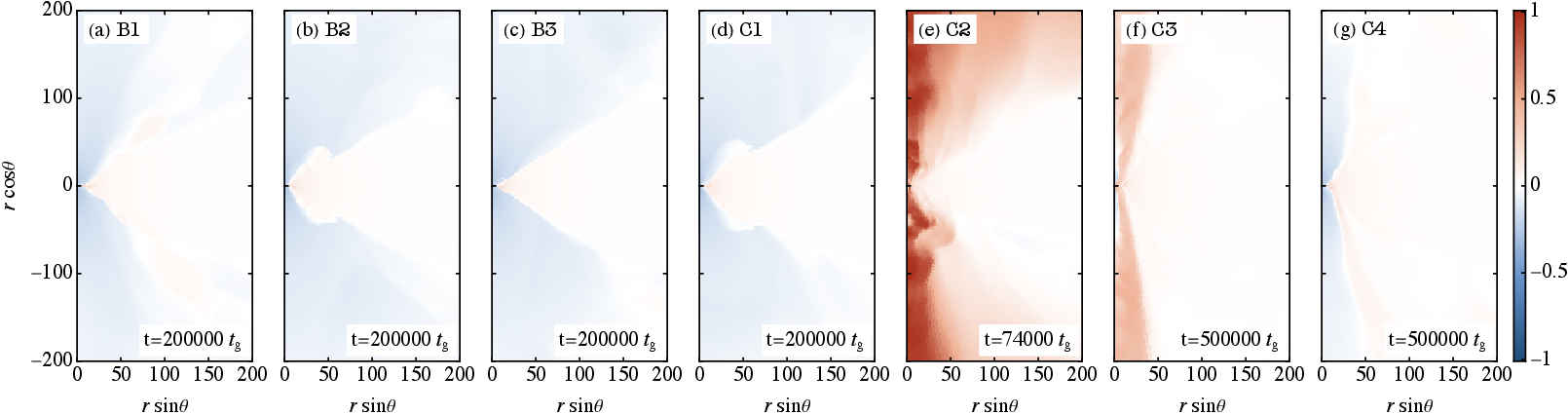}
\caption{For runs~\texttt{B1} to \texttt{C4},
the radial velocity profile at the end of each simulation.}
\label{fig:BC_vr}
\end{figure*}

Figure~\ref{fig:B12_ts} shows the time series of the density-averaged inverse plasma beta and the normalized magnetic flux at the horizon.
Comparing the growth phase of each curve of $\abra{\betamag^{-1}}_\rho$ reveals that smaller values of $|\aBH|$ lead to weaker dynamo efficiency, but the difference is not significant.
However, the saturated values of $\abra{\betamag^{-1}}_\rho$ are similar regardless of the different black hole spins.
Among the four runs plotted, only \run{A1} launches collimated outflows around the poles, whereas the other runs have lower accumulated magnetic fluxes, and no clear polar outflow is seen during the simulated time;
see panels~(a) to (c) in Figure~\ref{fig:BC_vr}.

\begin{figure*}
\centering
\includegraphics[width=1\textwidth]{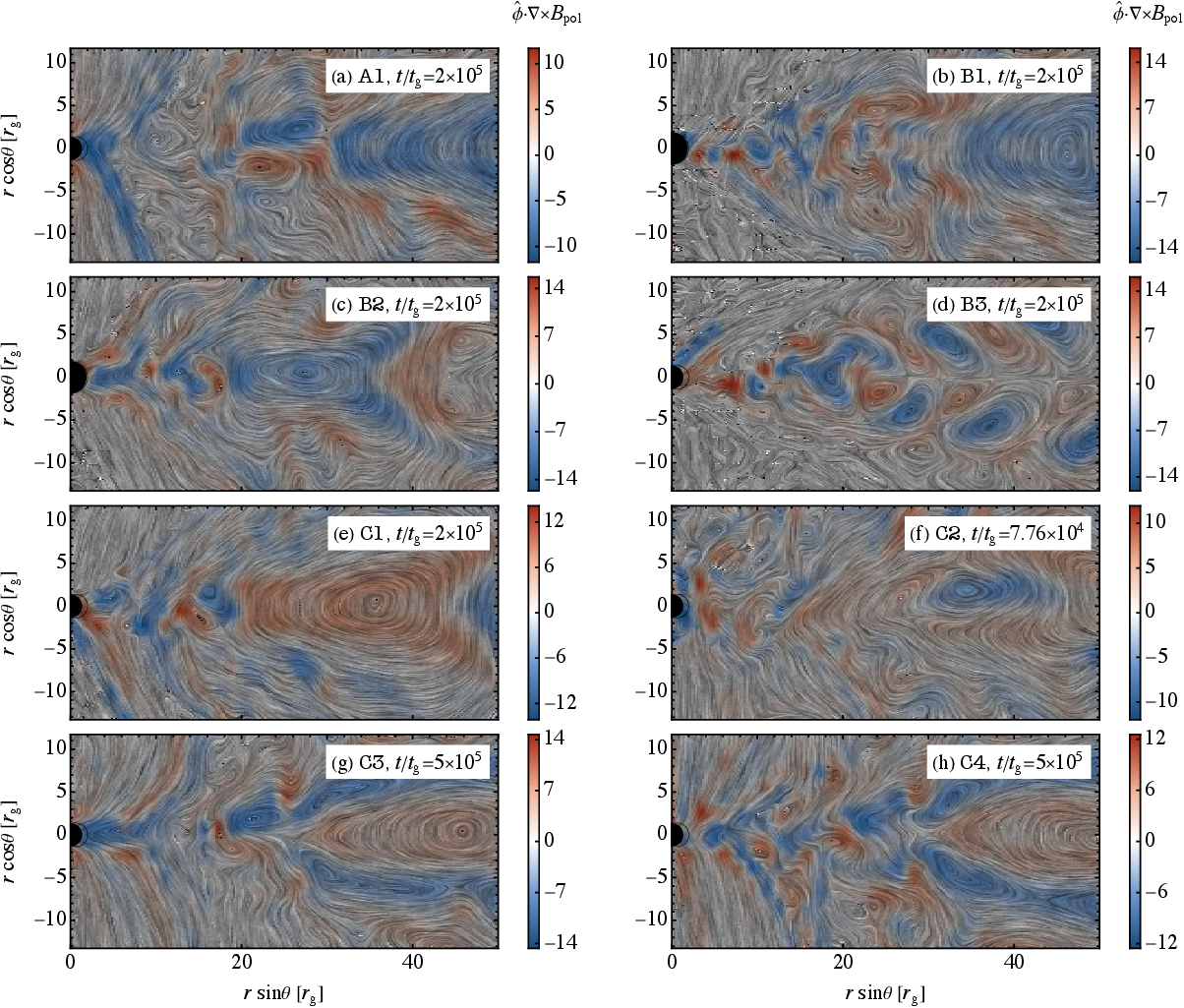}
\caption{Similar to Figure~\ref{fig:A1_Bpol},
but for runs~\texttt{A1} to \texttt{C4} at the end of each simulation.
Panel (a) reproduces Figure~\ref{fig:A1_Bpol}(b) as a reference.}
\label{fig:BC_Bpol}
\end{figure*}

In Figure~\ref{fig:BC_Bpol}(b) to (d), we plot the profile of the poloidal magnetic fields at the end of each simulation.
While the spatial scales of the magnetic fields remain similar across the runs,
\run{B3} shows clearly a multi-loop configuration with alternating polarity that resembles the Newtonian cases \citep{HoggReynolds2016,Zhou2024}.
A plausible explanation is that in the retrograde-spin case, the ISCO lies at a much larger radius.
Consequently, the stably rotating portion of the disk does not extend into the strongly relativistic region near the horizon,
and frame-dragging effects of a rotating black hole do not couple to the dynamo much, which makes the retrograde case appear more similar to the Newtonian cases.
 
\begin{figure*}
\centering
\includegraphics[width=0.9\textwidth]{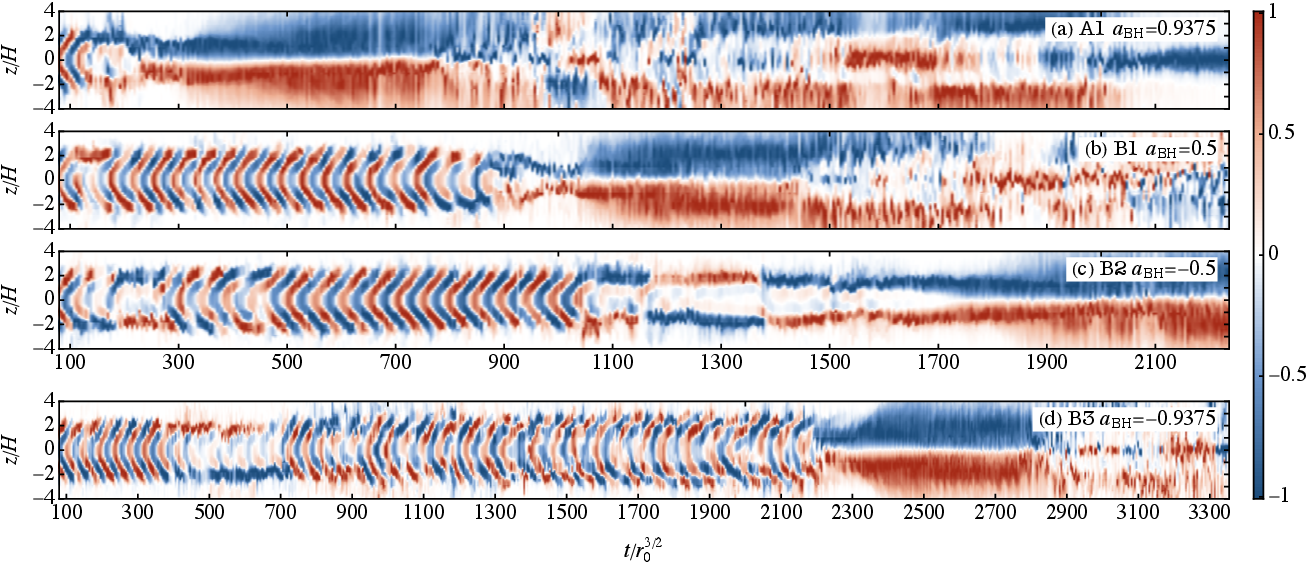}
\caption{Space-time diagrams for the $B_\phi$ polarity at $r=20\rg$ with varying black hole spin $a_\text{BH}$.
Panel(a) reproduces Figure~\ref{fig:A1_butterfly}(b), plotted here for reference.
Notice that \run{B3} spans a longer time range.
}
\label{fig:B12_butterfly}
\end{figure*}

Figure~\ref{fig:B12_butterfly} represents the space-time diagrams of $B_\phi$ normalized by its maximal strength at each fixed time for runs~\texttt{A1} and \texttt{B1} to \texttt{B3}.
The comparison suggests an interesting correlation between the persistence of the butterfly patterns and the black hole spin:
With more positive black hole spin, the butterfly pattern lasts for shorter and quickly transits to non-periodic or sporadically periodic states,
whereas for nearly maximally negative spin (\run{C3}), the cyclic pattern can last for more than $2000$ orbital time scales.
Again, the rapid retrograde-spin case appears the most similar to the persistent cycle periods shown in Newtonian simulations \citep[e.g.,][]{Gressel2010,BaiStone2013,HoggReynolds2018,Zhou2024},
reflecting the possibility that GR effects may intercept dynamo periodicities.

\subsection{Varying $\alphaSS$}

\begin{figure*}
\centering
\includegraphics[width=0.9\textwidth]{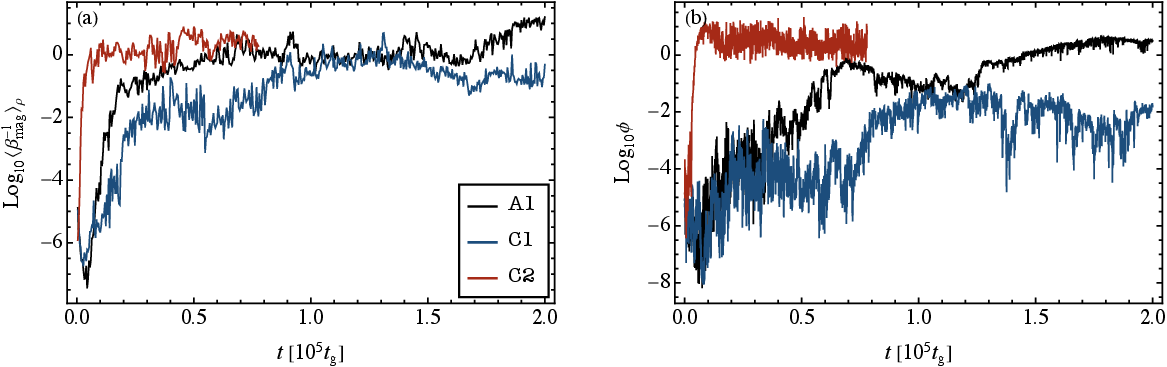}
\caption{With varying $\alphaSS$,
the time series of 
(a) the density-weighted averages of $\betamag^{-1}$, and
(b) normalized magnetic flux at the horizon.}
\label{fig:C12_ts}
\end{figure*}

We next examine the effect of changing the parameter $\alphaSS$,
thereby adjusting the strength of the turbulence and the dynamo.
For runs~\texttt{C1} and \texttt{C2},
while the poloidal field configurations do not show a qualitative difference compared to \run{A1} in Figure~\ref{fig:BC_Bpol}, both the growth rate and the saturation value of magnetization correlate positively with $\alphaSS$, as shown in Figure~\ref{fig:C12_ts}.
In particular, for \run{C1} ($\alphaSS=0.01$),
insufficient magnetic flux is accumulated on the event horizon [see panel~(b)] and hence we have not observed any polar outflow.
However, \run{C2} ($\alphaSS=0.1$) reaches $\phi=\mathcal{O}(5)$ at an early time and launches less collimated outflows compared to \run{A1}; see panels~(d) and (e) of Figure~\ref{fig:BC_vr}.
The comparison demonstrates that the results of the simulation depend sensitively on the input parameter $\alphaSS$, which is not known \textit{a priori}.
It is therefore desirable to build a model in which $\alphaSS$ can be deduced from other flow properties, and this is done in the next subsection.

\subsection{Towards a fully dynamical model}
\label{sec:dynamical}

Theories and numerical simulations have revealed that the turbulence vigorousness driven by MRI is directly correlated to the plasma beta.
However, the discovered scaling law between $\alphaSS$ and $\betamag$ is not universal. It depends on the initial value of $\betamag$ at the disk mid-plane in a shearing box.
For weakly magnetized MRI characterized by $\betamag(t=0,z=0)\gtrsim100$,
an empirical relation is found by \cite{Blackman+2008} from a large set of shearing-box simulations that $\alphaSS\simeq 0.1\betatot^{-1}$,
where $\betatot$ is the saturated plasma beta that includes both small- and large-scale magnetic fields.
\cite{Porth+2019} has also found a similar relation in radiatively inefficient accretion flows (RIAF) simulations across different GRMHD codes.
For mildly magnetized MRI with $\betamag(t=0,z=0)\lesssim100$, the saturated $\alphaSS$ is instead found to be correlated with the initial plasma beta, as $\alphaSS\simeq 10\betamag^{-1/2}(t=0,z=0)$ \citep[e.g.,][]{BegelmanArmitage2023}.

Our model belongs to the weak MRI limit, and we shall modify the relation of \cite{Blackman+2008} to connect $\alphaSS$ to $\betamag$, making use of the helicity balance relation~(\ref{eqn:helicity_balance}).
Clearly, we have $b^2/l\simeq B^2/3\alphaSS \hh$,
so that $b^2/B^2\simeq 1/\sqrt{3\alphaSS}$,
and $\betatot^{-1}\simeq \betamag^{-1}/\sqrt{3\alphaSS}$,
assuming that the total magnetic pressure is dominated by the small-scale fields.
The relation discovered in \cite{Blackman+2008} then yields
\beq
\alphaSS\simeq\left(300\betamag^2\right)^{-1/3}\simeq 0.15\betamag^{-2/3}.
\label{eqn:ass_dynamical}
\eeq
Such a relation allows for a fully dynamical subgrid model without free parameters and is implemented in runs~\texttt{C3} and \texttt{C4}.

\begin{figure*}
\centering
\includegraphics[width=0.8\textwidth]{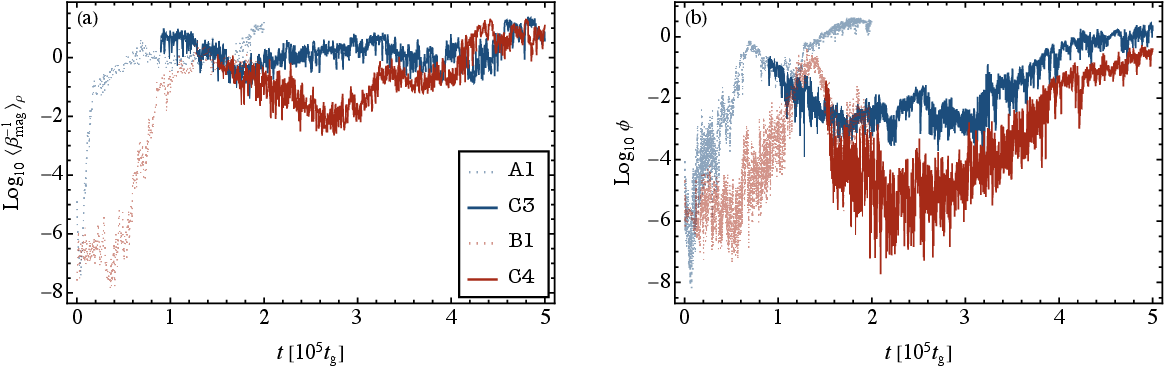}
\caption{For runs~\texttt{C3} and \texttt{C4}, time series of (a) the density-weighted averaged plasma beta and (b) normalized magnetic flux at the horizon.
Note that the runs restart from the $t=89200\tg$ and the $t=149600\tg$ snapshots of runs~\texttt{C3} and \texttt{C4}, respectively.
The dotted curves show the runs~\texttt{A1} and \texttt{B1} for references.
}
\label{fig:C34_ts}
\end{figure*}

Run~\texttt{C3} is the dynamical-$\alphaSS$ counterpart of \run{A1}.
To avoid the very long evolution time caused by the initially large $\betamag$ and small $\alphaSS$,
we restart from the $t=89200\tg$ snapshot of \run{A1} using the new prescription~(\ref{eqn:ass_dynamical}).
Similarly, \run{C4} restarts from \run{B1} at $t=149600\tg$.
As shown in Figures~\ref{fig:BC_Bpol} and  \ref{fig:C34_ts},
a dynamical $\alphaSS$ can maintain similar magnetization, magnetic flux at the horizon, and poloidal magnetic fields compared to the corresponding fixed $\alphaSS=0.02$ runs.

\begin{figure}
\centering
\includegraphics[width=0.45\textwidth]{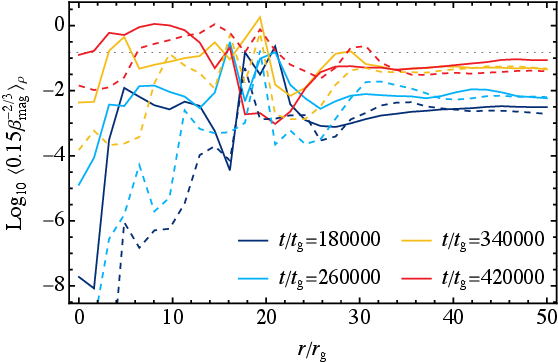}
\caption{Evolution of the radial profiles of $\alphaSS$ in runs~\texttt{C3} (solid) and \texttt{C4} (dashed).
Different color indicates different simulation times.
The dotted horizontal line indicates the threshold above which the dynamo is shut down, $\alphaSS(\betamag=1)=0.15$.
}
\label{fig:C34_ass}
\end{figure}

In Figure~\ref{fig:C34_ass}, we demonstrate that such a dynamical $\alphaSS$ prescription leads to a stationary radial profile of $\alphaSS$ at $r\gtrsim30\rg$, at $\alphaSS\simeq 0.035$ to $0.05$,
roughly double the value obtained from a global 3D simulation in \cite{Duez+2025}.
For both runs~\texttt{C3} and {C4},
during the time interval $3.4\times 10^5\leq t/\tg \leq 4.2\times 10^5$ (orange and red curves),
the inner disk region $r\lesssim15\rg$ has no dynamo action since $\betamag<1$, but the magnetization keeps increasing,
which indicates that the magnetization is caused by advecting field lines from the outer disk.
The high magnetization in the inner disk and the normalized magnetic flux greater than unity eventually lead to polar outflows in \run{C3} [see Figure~\ref{fig:BC_vr}(f)], similar to those shown in Figure~\ref{fig:A1_snaps}(f), with a Lorentz factor $\simeq 1.3$.
For \run{C4}, however, with reduced black hole spin, no sign of a polar outflow is seen,
similar to the case of \texttt{B1};
see Figure~\ref{fig:BC_vr}(g).

\section{Conclusion and discussion}
\label{sec:conclusion}

In this work, we have derived and implemented a subgrid helical mean-field dynamo model with dynamical quenching in resistive GRMHD simulations of geometrically thin accretion disks around a black hole.
The model requires only one minimal assumption,
namely that the disk is in a statistically steady state.
Other than that, our fully self-consistent model determines the dynamo-active region and quenching process according to MHD physical quantities with no \textit{ad hoc} prescriptions.
The unresolved turbulence strength and scales are determined by the input parameter $\alphaSS$, but we have also demonstrated an implementation where $\alphaSS$ can be locally determined by the magnetization.
The dynamo is quenched according to the helicity conservation constraint, without any pre-determined saturation level.

We have explored the consequences of the subgrid dynamo in geometrically thin disks.
In the strong dynamo ($\alphaSS=0.1$) or mild dynamo with rapid black hole spin ($\alphaSS=0.02$ and $\aBH=0.9375$) case,
the normalized magnetic flux at the horizon can reach $\gtrsim1$ values, and polar outflows with $\sigma<1$ and $\Gamma\simeq1.2$ are launched.
Otherwise, with lower $\alphaSS$ or $\aBH$, or retrograde black holes, no outflow is seen, albeit that the saturated density-weighted plasma beta remains similar.

For all the runs, we can reproduce multi-loop poloidal fields and butterfly diagrams of the azimuthal field, but the rapid retrograde-spin case shows the most regular poloidal loops, and its quasi-period lasts for the longest time.
We suggest that the correlation between the black hole spin and the similarity to Newtonian cases might be an indication of the coupling between GR effects and large-scale dynamos.

With either fixed $\alphaSS$ or dynamical $\alphaSS$,
our subgrid dynamo can be readily implemented in two- or three-dimensional simulations to allow for self-consistent magnetic field amplification,
without the need for highly resolved MRI turbulence.
With the reduction of the resolution requirement, it also becomes available to perform very long-time co-evolution of accretion disks and magnetic fields that could have interesting implications on state transition in X-ray binaries and different AGN models.

We note that in our model, the requirement of a steady state is still quite restrictive and prevents our model from being used straightforwardly in simulations with changing-structure disks, such as precessing disks or transient states like tidal disruption events (TDEs) and binary neutron-star mergers.
In such cases, an estimated and prescribed mean density profile could be helpful, but calculating the mean density on-the-fly when running simulations will ultimately be necessary.

Our model is also limited to an isotropic dynamo mechanism, i.e.,
$e^\mu=\xi b^\mu+\eta j^\mu$ with both $\xi$ and $\eta$ being scalars.
Several works have suggested that MRI dynamo may have an anisotropic origin \citep{GresselPessah2015,Dhang+2024MRI,Jacquimin-Ide+2024,Mondal+2025}. Thus, it is plausible to promote both $\xi$ and $\eta$ to be tensors for a more realistic MRI model.

Except for the high black hole spin model, our disk dynamo produces weakly magnetized thin disks as expected in the high state of X-ray binaries and radio-quiet AGNs.
During the state transition of X-ray binaries, strongly magnetized and geometrically thin disks may appear \citep{Pariev+2003,BegelmanPringle2007,DexterBegelman2019,Mishra+2020,Dhang+2025}, which could result from some dynamo model in the strong MRI regime \citep{PessahPsaltis2005,Salvesen+2016}.

\section*{Acknowledgments}

We thank Indu K. Dihingia for sharing the thin-disk setup and useful discussions.
This research is supported by
the National Natural Science Foundation of China (grant
No. 12403020, 12203033, 12273022, 12511540053),
the National Key R\&D Program of China (grant No. 2023YFE0101200),
the China Postdoctoral Science Foundation (No. 2023M732251, 2022M712086, and BX20220207),
the Shanghai Municipality orientation program of Basic Research for
International Scientists (grant No. 22JC1410600),
and Qimeng Project at Shanghai Polytechnic University.
The numerical simulations in this work were carried out on the Siyuan Mark-I and the ARM platform clusters supported by the Center for High Performance Computing at Shanghai Jiao Tong University, the Astro cluster supported by Tsung-Dao Lee Institute, and the Hefei Advanced Computing Center.


\bibliographystyle{aasjournalv7}
\bibliography{refs}

\end{document}

%% file: macros.tex
\def\apjs{ApJS}
\def\aap{A \& Ap}

\def\apjl{ApJL}
\def\aap{A\&A}

\newcommand{\beq}{\begin{equation}}
\newcommand{\eeq}{\end{equation}}

\def\bar{\overline}
\newcommand{\abra}[1]{\left\langle{#1}\right\rangle}
\newcommand{\emf}{\mathcal{E}}

\newcommand{\del}{{\bm\nabla}}

\newcommand{\Rm}{\text{Rm}}

\newcommand{\hh}{H}
\newcommand{\alphaSS}{\alpha_\text{SS}}
\newcommand{\cs}{c_\text{s}}
\newcommand{\rg}{r_\text{g}}
\newcommand{\betamag}{\beta_\text{mag}}
\newcommand{\betatot}{\beta_\text{tot}}
\newcommand{\betacrit}{\beta_\text{crit}}
\newcommand{\tg}{t_\text{g}}
\newcommand{\run}[1]{run~\texttt{{#1}}}
\newcommand{\xik}{\xi_\text{k}}
\newcommand{\xid}{\xi_\text{d}}

\newcommand{\aBH}{a_\text{BH}}

%% file: runs.tex
\begin{tabular}{crrrr}
\hline
Run & $\alphaSS$ & $a$ & $\log_{10}\beta_\text{sat}$ & $\log_{10}\phi_\text{sat}$\\
\hline
\texttt{A0} & $0$ & 0.9375 & 7.0 & -8.3\\
\texttt{A1} & $0.02$ & 0.9375 & -0.7 & 0.5\\
\hline \texttt{B1} & $0.02$ & 0.5 & 0.2 & -2.9\\
\texttt{B2} & $0.02$ & -0.5 & 0.2 & -0.3\\
\texttt{B3} & $0.02$ & -0.9375 & 0.2 & -1.1\\
\hline \texttt{C1} & $0.01$ & 0.9375 & 0.7 & -1.9\\
\texttt{C2} & $0.1$ & 0.9375 & -0.5 & 0.4\\
\texttt{C3} & $0.15\betamag^{-2/3}$ & 0.9375 & -1.0 & 0.3\\
\texttt{C4} & $0.15\betamag^{-2/3}$ & 0.5 & -0.9 & -0.5\\
\hline
\end{tabular}